\documentclass{svproc}
\usepackage{url}

\usepackage{bm}       % bold math
\usepackage{amsmath}
\usepackage{amssymb}

\begin{document}
\mainmatter              % start of a contribution
\title{The problem of cluster separability\\ in relativistic few-body systems}
\titlerunning{Cluster Separability}  % abbreviated title (for running head)
\author{Wolfgang~Schweiger\inst{1} \and Nikita~Reichelt\inst{1}
\and William~H.~Klink\inst{2}}
\authorrunning{W. Schweiger et al.} % abbreviated author list (for running head)
%Craig Chambers, Kim B. Bruce, and Elisa Bertino}
%
\institute{Institute of Physics, University of Graz, A-8010 Graz, Austria,\\
\email{wolfgang.schweiger@uni-graz.at}
\and
Deptartment of Physics and Astronomy,University of Iowa, Iowa City, USA
}

\maketitle              % typeset the title of the contribution

\begin{abstract}
An appropriate framework for dealing with hadron structure and hadronic physics in the few-GeV energy range is relativistic quantum mechanics. The Bakamjian-Thomas construction provides a systematic procedure for implementing interactions in a relativistic invariant way. It leads, however, to problems with cluster separability. It has been known for some time, due to Sokolov's pioneering work, that mass operators with correct cluster properties can be obtained through a series of unitary transformations making use of so-called {\em packing operators}. In the present contribution we sketch an explicit construction of packing operators for three-particle systems consisting of distinguishable, spinless particles.
\end{abstract}
\section{Relativistic Quantum Mechanics}
By a {\em relativistic quantum mechanics} we mean a quantum theory for a finite number of particles  invariant under Poincar\'e transformations. Speaking more formally, one has to find a representation of all the Poincar\'e generators in terms of self-adjoint operators on an appropriate Hilbert space such that these operators satisfy the Poincar\'e algebra. For an interacting theory it is well known that at least three of the Poincar\'e generators have to contain interaction terms.  Depending on which of the Poincar\'e generators become interaction dependent, one distinguishes different forms of relativistic dynamics~\cite{Dirac:49}. For our purposes it  turns out to be most convenient to use the point-form of relativistic dynamics in which all the components of the four-momentum operator $P^\mu$ contain interactions, whereas the generators of Lorentz-transformations stay interaction free~\cite{Kli:2018}.

The only systematic procedure for implementing interactions in the Poincar\'e generators of an $N$-particle system such that the Poincar\'e algebra is preserved was suggested long ago by Bakamjian and Thomas~\cite{Bakamjian:53}. In the point form this procedure amounts to factorizing the four-momentum operator into a four-velocity operator and a mass operator and putting interaction terms into the mass operator:
\begin{equation}\label{eq:BT}
{P^\mu}={M} { V^\mu_{\mathrm{free}}}= ({ M_{\mathrm{free}}}+{ M_{\mathrm{int}}})  { V^\mu_{\mathrm{free}}}={ P^\mu_{\mathrm{free}}}+{ P^\mu_{\mathrm{int}}}\,.
\end{equation}
Since the mass operator is a Casimir operator of the Poincar\'e group, the constraints on the interaction terms that guarantee Poincar\'e invariance become simply that $M_{\mathrm{int}}$ should be a Lorentz scalar and that it should commute with $V^\mu_{\mathrm{free}}$, i.e. $[M_{\mathrm{int}}, V^\mu_{\mathrm{free}}]=0$.

A very convenient basis for representing Bakajian-Thomas (BT) type mass operators consists of velocity states~\cite{Kli:2018},
$
\vert \vec{v}; \vec{k}_1, \mu_1; \vec{k}_2, \mu_2; \dots; \vec{k}_N, \mu_N
\rangle$.
These specify the state of an $N$-particle system by its overall velocity $\vec{v}$, the particle momenta $\vec{k}_i$ in the rest frame of the system ($\sum_{i} \vec{k}_i = 0$) and the (canonical) spin projections $\mu_i$ of the individual particles. The overall velocity factors out in velocity-state matrix elements of BT-type mass operators, leading to the separation of overall and internal motion of the system.

\section{Cluster Separability and Packing Operators}
An important requirement for a quantum mechanical system, in addition to relativistic invariance, is {\em cluster separability}. Cluster separability roughly means that subsystems of a quantum mechanical system should behave independently, if they are sufficiently space-like separated. There are weaker and stronger notions of cluster separability depending, e.g., on whether it is demanded for the $S$-operator or the Poincar\'e generators~\cite{Kei:91}. Also the kind of convergence
of these quantities when letting the separation distance go to infinity plays an important role. We are interested in cluster separability of the Poincar\'e generators and, in particular, of the invariant mass operator. For our purposes it is useful to introduce a purely algebraic notion of cluster separability which we call {\em coupling constant separability}. Coupling constant separability means that, for a particular clustering, after separation of the clusters, operators behave as if there were no interaction between the clusters. It does not say much about the range of the interactions, but for the construction we are going to make it is a useful and reasonable concept.

Problems with cluster separability start to show up for interacting three-particle systems and are closely connected with how the two-particle subsystems are implemented in the three-particle Hilbert space. Let us assume for simplicity spinless, distinguishable particles and start with BT-type four-momentum operators for the two-particle subsystems,
$P^\mu_{ij}=M_{ij}\,V^\mu_{ij\,\mathrm{free}}$, $i,j=1,2,3$, $i\neq j$.
The third particle is then added by means of the usual tensor-product construction,
\begin{equation}
\tilde{P}^\mu_{ij|k}:=P^\mu_{ij}\otimes I_{k}+I_{ij}\otimes P^\mu_{k}\,.
\end{equation}
The individual four-momentum operators
$\tilde{P}^\mu_{ij|k}$ describe 2+1-body systems in a Poincar\'e invariant way and also exhibit coupling constant separability. One may now think of adding the four-momentum operators for the different clusterings, as in the non-relativistic case, to end up with a four momentum operator for a three-particle system with pairwise interactions,
$
\tilde{P}^\mu_{123}=\tilde{P}^\mu_{12|3}+\tilde{P}^\mu_{23|1}+\tilde{P}^\mu_{31|2}-2 {P}^\mu_{123\,\mathrm{free}}\, .
$
But the components of $\tilde{P}_{123}$ do not commute,
$[\tilde{P}^\mu_{123},\tilde{P}^\nu_{123}]{\neq} 0$ since
$[M_{ij\,\mathrm{int}},V^\mu_{j}]\neq0$. The three-particle system, described by the four-momentum operator $\tilde{P}^\mu_{123}$, is thus not relativistically invariant. The first step to overcome this problem is to factorize the four-momentum operators for the different clusterings again into a mass operator and a four-velocity operator:
\begin{equation}
\tilde{P}^\mu_{ij|k}=\tilde{M}_{ij|k}\, \tilde{V}^\mu_{ij|k}\quad\hbox{with} \quad\tilde{M}_{ij|k}^2=\tilde{P}_{ij|k}\cdot \tilde{P}_{ij|k}\, .
\end{equation}
The four-velocities $\tilde{V}^\mu_{ij|k}$ contain interactions and differ for different clusterings. The key observation is now that there exist unitary operators which relate the four-velocity operators, since they all have the same spectrum ($\mathbb{R}^3$). One can find, in particular, unitary operators $U_{ij|k}$ such that
\begin{equation}\label{eq:inter}
\tilde{V}^\mu_{ij|k}=U_{ij|k}\, V^\mu_{123\,\mathrm{free}}\, U_{ij|k}^\dag\, .
\end{equation}
Applied to the four-momentum operators $\tilde{P}^\mu_{ij|k}$ these unitary operators pack the interaction dependence of the four-velocity operators $\tilde{V}^\mu_{ij|k}$ into the mass operator $M_{ij|k}=U^\dag_{ij|k} \tilde{M}_{ij|k} U_{ij|k}$ such that $P^\mu_{ij|k}=U^\dag_{ij|k} \tilde{P}^\mu_{ij|k} U_{ij|k}=M_{ij|k} V^\mu_{123\,\mathrm{free}}$ is of BT-type. Therefore $U_{ij|k}$ were called {\em packing operators} by Sokolov in his seminal paper on the formal solution of the cluster problem~\cite{Sokolov:78}. The new four-momentum operator $P^\mu_{123\,\mathrm{BT}}=P^\mu_{ij|k}+P^\mu_{ij|k}+P^\mu_{ij|k}-2P^\mu_{123\,\mathrm{free}}=
M_{123\,\mathrm{BT}}V^\mu_{123\,\mathrm{free}}$ is now also of BT-type and thus provides  a relativistic invariant description of a three particle system with pairwise two-particle interactions;  but it still misses coupling-constant separability, as one can easily check. The way out is a further unitary transformation which involves the packing operators we have already introduced:
\begin{equation}
P^\mu_{123} =\left(\prod U_{ij|k}\right)\,P^\mu_{123\,\mathrm{BT}}\,\left(\prod U_{ij|k}\right)^\dag\, .
\end{equation}
If $U_{ij|k}\rightarrow 1$ for separations $ki|j$, $jk|i$ and $i|j|k$ and if $(\prod U_{ij|k})$ commutes with the generators of Lorentz transformations, it can be shown that such a {\em generalized BT construction} leads to relativistic invariance and coupling-constant separability of the resulting three-body model~\cite{Sokolov:78}. 

The procedure just outlined solves the cluster problem for three-body systems formally, but its practical applicability depends strongly on the capability to calculate the packing operators. The solution to this problem can also be found in Sokolov's paper~\cite{Sokolov:78}. The trick is to split the packing operator further into a product of unitary operators which depend on the corresponding two-particle mass operators in a way to be determined, $U_{ij|k}=W^\dag(M_{ij}) W(M_{ij\,\mathrm{free}})$. With this splitting one can rewrite Eq.~(\ref{eq:inter}) in the form
$
 W(M_{ij\,\mathrm{free}}) V^\mu_{123\,\mathrm{free}} W^\dag(M_{ij\,\mathrm{free}})=W(M_{ij})\tilde{V}^\mu_{ij|k}W^\dag(M_{ij})\, .
$
Since this equation should hold for any interaction, the right- and left-hand sides can be chosen to equal some simple four-velocity operator, for which $V_{ij\,\mathrm{free}}^\mu\otimes I_k$ is a good choice. In order to compute the action of $W$ it is then convenient to take bases in which matrix elements of $V^\mu_{123\,\mathrm{free}}$, $V_{ij\,\mathrm{free}}^\mu\otimes I_k$ and $\tilde{V}^\mu_{ij|k}$ can be calculated. This is the basis of (mixed) velocity eigenstates
$
|\vec{v}_{ij};\vec{\tilde{k}}_i,\vec{\tilde{k}}_j,\vec{p}_k\rangle=
|\vec{v}_{ij};\vec{\tilde{k}}_i,\vec{\tilde{k}}_j\rangle\otimes|\vec{p}_k\rangle
$
of $M_{ij|k\,\mathrm{free}}$ if one wants to calculate the action of $W(M_{ij\,\mathrm{free}})$ and corresponding eigenstates of $M_{ij|k}$ if one wants to calculate the action of $W(M_{ij})$. It turns out that the effect of these operators is mainly to give the two-particle subsystem $ij$ the velocity $\vec{v}_{ij|k}$ of the whole three-particle system. After some calculations one finds  that the whole effect of the packing operator $U_{ij|k}$ on the mass operator $\tilde{M}_{ij|k}$ is just the replacement of kinematical factors in the mixed velocity-state matrix elements, so that the transformed mass operator $U_{ij|k}\, \tilde{M}_{ij|k}\, U_{ij|k}$ attains BT-type structure~\cite{Kli:2018}:
\begin{eqnarray}
&\frac{1}{m_{ij}^{\prime\, 3/2} m_{ij}^{3/2}}
\, v^0_{ij}\,\delta^3(\vec{v}_{ij}^{\,\,\prime}-\vec{v}_{ij})&\nonumber \\ & \downarrow & \\ &
\frac{\sqrt{v_{ij}^{\,\,\prime} \cdot {v}_{123}}}{m_{123}^{\prime\, 3/2}}
\,\frac{\sqrt{v_{ij}^{} \cdot {v}_{123}}}{m_{123}^{3/2}}\, v^0_{123}\,\delta^3(\vec{v}_{123}^{\,\,\prime}-\vec{v}_{123})&\, .\nonumber
\end{eqnarray}
Here $m_{ij}$ and $m_{123}$ are the invariant masses of the non-interacting two-particle subsystem $ij$ and the non-interacting three-particle system, respectively, $v_{ij}$ and $v_{123}$ are the corresponding four-velocities.

\section{Concluding Remarks}
Knowing the packing operators and their action on the mass operators $\tilde{M}_{ij|k}$, we can construct a three-body mass operator $M_{123}=(\prod U_{ij|k})\, M_{123\, BT}\, (\prod U_{ij|k})^\dag$ for distinguishable, spinless particles that has correct cluster properties. The inclusion of spin should be straightforward. The treatment of identical particles is, however, more intricate, because the simple product $(\prod U_{ij|k})$ has to be replaced by some kind of symmetrized product to preserve the symmetry under exchange of particles. Although the formal solution of the cluster problem and explicit expressions for the packing operators have been known for quite some time~\cite{Sokolov:78,Kei:91}, cluster separability has always been  neglected in practical applications. This may well be  justified for weakly bound nuclear systems~\cite{Keister:2011ie}, but has to be thoroughly investigated for strong binding.  With our results we are now, at least in the simple case of spinless, distinguishable particles,  in the position to  perform numerical studies for clarifying the precise role of cluster separability in relativistic  three-body systems. 


\end{document}